\documentclass[prd,amsfonts,amsmath,twocolumn,nofootinbib]{revtex4}

\usepackage{color}
\usepackage{graphicx} 
\usepackage{epstopdf}%
\usepackage{mathptmx}   

\usepackage{multirow}

\newcommand {\be} {\begin{equation}} 
\newcommand {\ba}{\begin{eqnarray}} 
\newcommand {\ee} {\end{equation}} 
\newcommand{\ea} {\end{eqnarray}}

\renewcommand{\Im}{{\rm Im\,}} 
\renewcommand{\Re}{{\rm Re\,}}

\begin{document}

\title{Modification of electromagnetic structure functions for the $\gamma Z$-box diagram} 

\author{Benjamin C. Rislow}

\author{Carl E.\ Carlson}

\affiliation{Department of Physics, College of William and Mary, Williamsburg, VA 23187, USA}

\date{June 27, 2013}

\begin{abstract}

The $\gamma Z$-box diagram for parity violating elastic $e$-$p$ scattering has recently undergone a thorough analysis by several research groups.  Though all now agree on the analytic form of the diagram, the numerical results differ due to the treatment of the structure functions, $F_{1,2,3}^{\gamma Z}(x,Q^2)$.  Currently, $F_{1,2,3}^{\gamma Z}(x,Q^2)$ at low $Q^2$ and $W^2$ must be approximated through the modification of existing fits to electromagnetic structure function data.  We motivate and describe the modification used to obtain $F_{1,2}^{\gamma Z}(x,Q^2)$ in our previous work.  We also describe an alternative modification and compare the result to our original calculation.  Finally, we present a new modification procedure to acquire $F_3^{\gamma Z}(x,Q^2)$ in the resonance region and calculate the axial contribution to the $\gamma Z$-box diagram.  Details of these modifications will illuminate where discrepancies between the groups arise and where future improvements can be made.  

\end{abstract}

\maketitle


\section{Introduction}			\label{sec:one}


Parity violating $e$-$p$ scattering experiments performed at momentum transfers away from the $Z$-pole are used to test the Standard Model prediction of the running of $\text{sin}^2\theta_W$.  The $Q_{weak}$ experiment at Jefferson Lab~\cite{Rajotte:2011nn} aims to perform a 0.3\% measurement of $\text{sin}^2\theta_W$ at a momentum transfer of $Q^2=0.026 \text{ GeV}^2$.  To obtain this desired precision, all radiative corrections must be known to an even higher precision.  Up to one loop order, the weak charge of the proton at zero momentum transfer is given by~\cite{Erler:2003yk}
\begin{align}
						\label{eq:corrections}
Q_W^p &= \left( 1+\Delta\rho + \Delta_e \right) \left(1-4\sin^2\theta_W(0) + \Delta'_e \right)
\nonumber\\
& + \square_{WW} + \square_{ZZ} + \Re\square_{\gamma Z} .
\end{align}

Here, $\Delta_e$ and $\Delta_e'$ are electron vertex corrections, $\Delta\rho$ is the W and Z mass renormalization, and $1-4\sin^2\theta_W(0)$ is the one loop value of the weak mixing angle evaluated at $Q^2=0$.  The $WW$ and $ZZ$ box diagrams, $\square_{WW}$ and $\square_{ZZ}$, are dominated by large momentum exchange and can be calculated using perturbative QCD.  A different technique is required to calculate the $\gamma Z$-box diagram due to low $Q^2$ contributions.  Gorchtein and Horowitz~\cite{Gorchtein:2008px} used a dispersion relation to evaluate the $\gamma Z$-box diagram at zero momentum transfer and obtained a result that was larger than expected~\cite{Erler:2003yk}.  Sibirtsev \textit{et al.}~\cite{Sibirtsev:2010zg} used the same technique and found an analytic result that was greater by a factor of 2.  This discrepancy inspired a third calculation~\cite{Rislow:2010vi} that agreed with the Sibirtsev \textit{et al.} result.  After reevaluating their work, Gorchtein \textit{et al.}~\cite{Gorchtein:2011mz} confirmed the factor of 2.  All three groups now agree on the analytic form of the $\gamma Z$-box.  The imaginary vector portion is
\begin{align}
\label{imdelta3}
\Im \square^V_{\gamma Z}(E) &= \frac{\alpha_{em}}{(2ME)^2}
\int^s_{W^2_\pi} dW^2 
				\nonumber \\
& \times \int^{Q^2_{max}}_0 dQ^2 \frac{ F^{\gamma Z}_1(x,Q^2)+AF^{\gamma Z}_2(x,Q^2)}{1 + Q^2/M^2_Z }	\,,
\end{align}
where
\be
\label{A}
A= \frac{(2ME)^2-2ME(W^2-M^2+Q^2)-M^2Q^2}{Q^2(W^2-M^2+Q^2)}.
\ee
In the above equations $M$ is the mass of the proton, $E$ is the lab energy of the incoming electron, $s=M^2+2ME$, $W_\pi^2=(M+m_\pi)^2$, $m_\pi$ is the mass of the pion, and $Q_{max}^2=(s-M^2)(s-W^2)/s$.  The fine structure constant $\alpha_{em}(Q^2=0)$ is used because the integral receives most of its support from low $Q^2$.  The dispersion relation that relates $\Im\square_{\gamma Z}^V$ to $\Re\square_{\gamma Z}^V$ is
\be
\label{redelta}
\Re \square^V_{\gamma Z}(E)=\frac{2E}{\pi} \int^\infty_{\nu_\pi}\frac{dE'}{E'^2-E^2}\Im \square^V_{\gamma Z}(E') 
\ee
where $\nu_\pi=(W^2_\pi-M^2)/2M$.

The $Q_{weak}$ experiment ran at an incoming electron energy of E = $1.165$ GeV.  Table~\ref{vec} shows the numerical $\Re\square_{\gamma Z}^V$ results obtained by each group at this energy.  The differences occur because of the models used for the $F_{1,2}^{\gamma Z}$ structure functions.  Currently, there are no data for these structure functions at low $Q^2$ and $W^2$ and each group performed calculations using their own modifications to electromagnetic structure functions.  The PVDIS experiment~\cite{Wang:2013uza} at Jefferson Lab has several data points for the deuteron's $F_{1,2,3}^{\gamma Z}$ in the resonance region.  These data will be insufficient to produce a model-independent fit, but provide a first step in testing the validity of the modifications~\cite{Carlson:2012yi}.  

\begin{table}[h]

\caption{$\Re\square_{\gamma Z}^V\times 10^3$ evaluated at E = $1.165$ GeV.}
\begin{tabular}{|l l|}
\hline
Sibirtsev \textit{et al.}~\cite{Sibirtsev:2010zg} & $4.7_{-0.4}^{+1.1}$ \\ 
Rislow and Carlson~\cite{Rislow:2010vi} & $5.7 \pm 0.9$ \\ 
Gorchtein \textit{et al.}~\cite{Gorchtein:2011mz}& $5.4 \pm 2.0$  \\ 
 Hall \textit{et al.}~\cite{Hall:2013hta}   & $5.60 \pm 0.36$ \\ \hline
\end{tabular}
\label{vec}
\end{table}

The axial contribution to the $\gamma Z$-box has also recently undergone analysis.  The axial contribution to $\Im\square_{\gamma Z}$ is
\begin{align}
\label{imdeltaax}
\Im \square^A_{\gamma Z}(E) &= \frac{1}{(2ME)^2}
\int^s_{M^2} dW^2 
				\nonumber \\
& \times \int^{Q^2_{max}}_0 dQ^2\alpha_{em}(Q^2)\frac{g^e_V(Q^2)}{g^e_A} \frac{B F^{\gamma Z}_3(x,Q^2)}{1+Q^2/M_Z^2}	\,,
\end{align}
where
\be
\label{B}
B= \frac{2ME}{W^2-M^2+Q^2}-\frac{1}{2}.
\ee
The weak couplings for the electron are given by $g^e_V=T_e^3-2Q_e\text{sin}^2\theta_W(Q^2)$, and $g^e_A=T_e^3$.  The axial integral receives support from high $Q^2$ and we allow both $\alpha_{em}$ and $\sin^2\theta_W$ to run.  The dispersion relation that relates $\Im\square_{\gamma Z}^A$ to $\Re\square_{\gamma Z}^A$ is
\be
\label{redelta}
\Re \square^A_{\gamma Z}(E)=\frac{2}{\pi} \int^\infty_{\nu_\pi}dE'\frac{E'}{E'^2-E^2}\Im \square^A_{\gamma Z}(E').
\ee

Blunden \textit{et al.}~\cite{Blunden:2011rd} obtained axial results of the same order of magnitude as $\Re\square_{\gamma Z}^V$.  Repeating a similar analysis we have also calculated $\Re\square_{\gamma Z}^A$.  The two results for the axial contribution at the $Q_{weak}$ energy are reported in Table~\ref{ax}.  As with the $\Re\square_{\gamma Z}^V$ calculation, differences between the axial results occur because of the structure function treatment. 

The goal of our paper is to describe our modifications to the electromagnetic structure functions.  In Section II we present the steps taken to obtain $F_{1,2}^{\gamma Z}$ in the resonance region.  We focus attention on this region since most of the support for the vector $\gamma Z$-box integral comes from low $Q^2$.  These steps were not described in detail in our previous work and will allow a more thorough assessment of our $\Re\square_{\gamma Z}^V$ calculation.  In Section III we describe an alternative modification for obtaining $F_{1,2}^{\gamma Z}$ in the resonance region.  This modification is similar to the one used by Gorchtein \textit{et al.}~\cite{Gorchtein:2011mz} and the close agreement to our original $\Re\square_{\gamma Z}^V$ result suggests both modifications are equally valid, at least for the $Q_{weak}$ kinematics.   In Section IV we present our calculation of $F_3^{\gamma Z}$ in the resonance region that parallels the analysis of Section II.  We compare $F_3^{\gamma Z}$ and $\Re\square_{\gamma Z}^A$ values to those obtained by Blunden \textit{et al.}  Concluding remarks are contained in Section V.

\begin{table}

\caption{$\Re\square_{\gamma Z}^A\times 10^3$ evaluated at E = $1.165$ GeV.}
\begin{tabular}{|l l|}
\hline
Blunden \textit{et al.}~\cite{Blunden:2011rd} & $3.7\pm 0.4$ \\ 
This Work & $4.0\pm 0.5$  \\ \hline
\end{tabular}
\label{ax}
\end{table}


\section{Modification of the Structure Functions $\bf{F_{1,2}^{\gamma \gamma}(x,Q^2)\rightarrow F_{1,2}^{\gamma Z}(x,Q^2)}$ in the Resonance Region.}
\label{sec:f1f2mod}


In our previous work we modified the Christy and Bosted fit to electromagnetic data~\cite{Christy:2007ve} in the the resonance region $Q^2$ $< 8$ GeV$^2$ and W $< 2.5$ GeV.  Their fits for $F_1^{\gamma \gamma}$, $\sigma_T$, and $\sigma_L$ account for the contributions of seven resonances as well as a smooth background.  Their description and computer code for their fit allowed us to separately modify the resonances and the background. 

To obtain the resonance part of $F_1^{\gamma \gamma}$, Christy and Bosted sum the contribution of each resonance, $F_1^{\gamma \gamma}|_{res}$.  The resonance part of $F_1^{\gamma Z}$ can be calculated by modifying the summation by the insertion of corrective prefactors:
\be
\label{Fsum}
F_1^{\gamma Z}=\sum_{res}C_{res}\times F_1^{\gamma \gamma}|_{res}.
\ee
The prefactors are simply a ratio of structure functions for each of the resonances,
\be
\label{coef}
C_{res}=\frac{F_1^{\gamma Z}}{F_1^{\gamma \gamma}}\bigg|_{res}.
\ee
We next convert $C_{res}$ into a ratio of helicity amplitudes.  Following the normalization of the Particle Data Group~\cite{Nakamura:2010zzi}, the resonant parts of these structure functions can be expressed as a product of the polarization vector, $\epsilon_+^\mu=1/\sqrt2 (0,-1,-i,0)$, and hadronic tensors:
\begin{align}
\label{f1structure}
F_1^{\gamma \gamma (\gamma Z)}\Big|_{res} &=\epsilon^{\mu*}_+\epsilon_+^{\nu}W_{\mu \nu}^{\gamma \gamma (\gamma Z)}
\nonumber\\
&=(2)\sum_{\lambda}\int \text{d$^4$z}e^{iqz}\big<N,s\big|\epsilon_+^*\cdot J^{\gamma (Z,V)\dagger}(z)\big|res,\lambda\big>
\nonumber\\
&\times\big<res,\lambda\big|\epsilon_+\cdot J^{\gamma}(0)\big|N,s\big>,
\end{align}
where $N$ is a nucleon, $\lambda$ and $s$ are the spin projections of the resonance and nucleon, respectively, and $\gamma$ (Z,V) is the electromagnetic (neutral vector) current.  The factor of 2 is present in $\gamma Z$-exchange to account for the different orderings.

The above amplitudes can be evaluated by considering $\epsilon_+ \cdot J$ as a quark operator embedded between $SU(6)$ wave function representations of the nucleon and resonances~\cite{Close:1979bt}.  This operator ignores the spatial wave functions, $\psi$, and acts only on the flavor, $\phi$, and spin, $\chi$, wave functions.  Because the colorless portion of the total hadronic wave function is symmetric, we are free to operate only on the third components of $\phi$ and $\chi$ and multiply the result by three.  The amplitude can be expressed as
\begin{align}
\big<res&,\lambda\big|\epsilon_+\cdot  J^{\gamma (Z,V)}\big|N,s\big>
\nonumber \\
&=3\langle \psi_{res}\phi_{res}\chi_{\lambda}\big| e_q^{(3)}\big(g_V^{q (3)}\big)\bar u_{k',\lambda'}\epsilon_+ \cdot \gamma u_{k,s'}\big|\psi_N\phi_N\chi_s\rangle,
\end{align}
where $k$ ($k'$) and $s'$ ($\lambda'$) are the initial (final) momentum and spin projection for the struck quark.  The superscript (3) over the quark electromagnetic and weak vector couplings, $e_q$ and $g_V^q$, indicates that the operators are acting only on the third quark.  

Using unit normalized quark spinors, 
\be
\label{spinor}
u_{p,s}=\sqrt{\frac{E+m_q}{2m_q}}\begin{pmatrix} \xi_s \\ \frac{\vec\sigma\cdot \vec p}{2m_q}\xi_s \end{pmatrix},
\ee
and choosing a frame where the gauge boson is propagating in the z-direction, the current reduces to
\be
\label{vectorcurrent}
\bar u_{k',\lambda'}\epsilon_+ \cdot \gamma u_{k,s'}=\frac{\sqrt{2}}{2m_q}\xi^\dagger_{\lambda'}[P_++q_zS_+]\xi_{s'},
\ee
where $m_q$ is the constituent quark mass, $P_+=k_1+ik_2$, $S_+=1/2(\sigma_1+i\sigma_2)$, $q_z$ is the momentum of the boson, and $\xi_s$ are the usual two spinors.  The Wigner-Eckart Theorem allows us to calculate a matrix element of $P_+$ as a constant times a matrix element of $L_+$.

After absorbing the spatial and momentum information, as well as the quark mass coefficient, into parameters A and B, Eq.(\ref{f1structure}) becomes
\begin{align}
F_1^{\gamma \gamma (\gamma Z)}&\Big|_{res} = 
\nonumber\\
& 3\langle \psi_N\phi_N\chi_s\big|e_q^{(3)}\big(2g_V^{q (3)}\big)[AL_++BS_+]^\dagger\big|\psi_{res}\phi_{res}\chi_\lambda\rangle
\nonumber\\
&\times 3\langle \psi_{res}\phi_{res}\chi_\lambda\big|e_q^{(3)}[AL_++BS_+]\big|\psi_N\phi_N\chi_s\rangle.
\end{align}
In terms of helicity amplitudes,
\begin{align}
\label{f1helicity}
F_1^{\gamma \gamma (\gamma Z)}\Big|_{res} &= A_{\lambda}^{\gamma}(2A_{\lambda}^{Z})
\nonumber \\
&\times A_{\lambda}^{\gamma}.
\end{align}
where the helicity amplitudes are given by
\begin{align}
\label{f1helicity}
A_{\lambda}^{\gamma}(2A_{\lambda}^{Z})=3\langle \psi_N\phi_N\chi_s\big|e_q^{(3)}\big(2g_V^{q (3)}\big)[AL_++BS_+]^\dagger\big|\psi_{res}\phi_{res}\chi_\lambda\rangle,
\end{align}
and $\lambda$ is the spin projection of the resonance along the direction of the gauge boson momentum, $\gamma (Z)$ is the exchanged boson.

The prefactor can now be expressed as
\begin{align}  
C_{res}&=2\frac{\sum_\lambda A_{\lambda}^{\gamma}A_{\lambda}^{Z}}{\sum_\lambda (A_{\lambda}^{\gamma})^2}.
\end{align}

In general, to calculate these amplitudes we operated the Hamiltonian on the $SU(6)$ spatial ($\psi$), flavor ($\phi$), and spin ($\chi$) wave functions of protons and resonances described by Close~\cite{Close:1979bt}.  As examples, the proton and $D_{13}(1520)$ resonance are members of the $(2^8,56)$ and $(2^8,70)$ multiplets respectively and can be written as
\be
\label{proton}
|2^8,56\rangle=\frac{1}{\sqrt{2}}\psi^S_{L=0,L_Z=0}\left(\phi^{M,S}\chi^{M,S}_{S_Z=\pm1/2}+\phi^{M,A}\chi^{M,A}_{S_Z=\pm1/2}\right)
\ee
\begin{align}
\label{D13}
|2^8,70\rangle&=\sum_{J_Z=S_Z+L_Z}\langle J = 3/2\text{ } J_Z|L L_Z, S S_Z\rangle
		\nonumber\\
& \times
\frac{1}{2}\Big[\psi^{M,S}_{L L_Z}\left(\phi^{M,S}\chi^{M,S}_{S_Z}-\phi^{M,A}\chi^{M,S}_{S_Z}\right)
\nonumber \\
& +\psi^{M,A}_{L L_Z}\left(\phi^{M,S}\chi^{M,A}_{S_Z}+\phi^{M,A}\chi^{M,S}_{S_Z}\right)\Big].
\end{align}
M,(A)S indicates a wave function with two elements that are (anti)symmetric.

Inserting the Hamiltonian into the proton to $D_{13}(1520)$ helicity amplitudes gives
\begin{align}
\label{A12}
A^{\gamma(Z)}_{\lambda=1/2}&=3\times e_q^{(3)}\big(g_V^{q (3)}\big)\langle \psi_{res}\phi_{res}\chi_{+1/2}\big|[AL_++BS_+]\big|\psi_N\phi_N\chi_s\rangle
\nonumber \\
&=\frac{1}{\sqrt{6}}\Big(-A_{10}\left[e_u(g^u_V)-e_d(g^d_V)\right]
\nonumber \\
&-\sqrt{2}B_{10}\left[\frac{5}{3}e_u(g^u_V)+\frac{1}{3}e_dg^d_V)\right]\Big)
\end{align}
and
\begin{align}
\label{A32}
A^{\gamma(Z)}_{\lambda=3/2}&=3\times e_q^{(3)}\big(g_V^{q (3)}\big)\langle \psi_{res}\phi_{res}\chi_{+3/2}\big|[AL_++BS_+]\big|\psi_N\phi_N\chi_s\rangle
\nonumber \\
&=-\frac{1}{\sqrt{2}}A_{10}\left[e_u(g^u_V)-e_d(g^d_V)\right],
\end{align}
The subscripts of $A_{10}$ and $B_{10}$ indicate the angular momentum dependence of the resonance's wave function.

Obtaining $A_{10}$ and $B_{10}$ without relying on hadronic wave function requires additional phenomenological information.  Data for both of the $D_{13}(1520)$ and $F_{15}(1680)$ resonances~\cite{Brodsky:1981kj,Carlson:1985mm} show that the polarization ratio
\be
\label{polar}
A = \frac{ |A_{1/2}^\gamma|^2 - |A_{3/2}^\gamma|^2 }
	{ |A_{1/2}^\gamma|^2 + |A_{3/2}^\gamma|^2 }
\ee
is close to $-1$ for photoproduction, and approaches $+1$ at higher $Q^2$ as the $A_{1/2}^\gamma$ amplitude dominates (in accord with perturbative QCD).  Looking at the expressions for the $D_{13}(1520)$, we conclude that 
\be
A_{10}(Q^2=0) = - \sqrt{2} B_{10}(Q^2=0)
\ee
and expecting $A_{1/2}^\gamma$ to dominate by a power of $Q^2$ at high $Q^2$,  we choose a form with the correct limits
\be
\label{ABrelation}
\frac{A_{10}(Q^2)}{B_{10}(Q^2)} = - \sqrt{2}\,  f_1(Q^2) 
	= - \sqrt{2} \frac{1}{1 + Q^2/\Lambda_1^2}  \,.
\ee
We can now express $A_{10}$ in terms of $f_1$ and $B_{10}$.  Substituting this new value of $A_{10}$ into Eqs.~(\ref{A12}) and (\ref{A32}) leads to the prefactor of $D_{13}(1520)$:
\be
C_{D_{13}} = \frac{ (\frac{1}{3} - f_1)(1- f_1) + 3 f_1^2 }
	{ (1- f_1)^2 + 3 f_1^2 }  +  Q_W^{p,LO}	\,,
\ee
where $Q_W^{p,LO}=1-4\sin^2\theta_W(0)$.  A parallel analysis gives
\be
C_{F_{15}} = \frac{ \frac{2}{3} (1- f_2)}
	{ (1- f_2)^2 + 2 f_2^2 }  +  Q_W^{p,LO}	\,.
\ee
We used $\Lambda_1^2 = \Lambda_2^2 = 0.2$ GeV$^2$ in~\cite{Rislow:2010vi}.  As a check, we can compare our fits constructed using Close's analysis with amplitude fits from Mainz (MAID)~\cite{Tiator:2011pw}.  Better agreement can be obtained by setting $\Lambda_1^2=0.256$ GeV$^2$ and $\Lambda_2^2=0.635$ GeV$^2$, but this more thorough analysis does not change the overall $\Re\square_{\gamma Z}^V$ result  by more than half a percent.

Table~\ref{amptable} summarizes the helicity amplitudes and prefactors for each resonance in the Christy and Bosted fit.  The Roper resonance, $P_{11}(1440)$, belongs to the same multiplet as the proton.  $AL_+$ does not contribute to the amplitude since both the Roper and proton have zero orbital angular momentum.  Consequently, the amplitude is only proportional to $B_{00}$ and the Roper prefactor is $Q^2$-independent.  For resonances with non-zero orbital angular momentum, $C_{res}$ is $Q^2$-dependent.  The two $S_{11}$ states belong to the same $SU(6)$ multiplet as the $D_{13}(1520)$, so $A_{10}$ and $B_{10}$ are the same for all three states, for valid SU(6) symmetry.  The $S_{11}$ states can mix.  We have written above the results for the unmixed case.  The unmixed $\gamma p$ amplitude for the $S_{11}(1650)$ is zero when the values of the quark charges are inserted;  this is the Moorhouse selection rule~\cite{Moorhouse:1966jn}.   If we neglect this amplitude also for the $Z$-boson case,  the amplitude listed for the $S_{11}(1535)$ gives a ratio
\be
C_{S_{11}} = \frac{ \frac{1}{3} + 2 f_1 }
	{ 1 + 2 f_1 }  +  Q_W^{p,LO}	\,.
\ee
Electroproduction of the $S_{11}(1650)$ occurs because of mixing with the bare $S_{11}(1535)$, and the above ratio is the same for both the $S_{11}$'s.  We have checked that including mixing makes little numerical difference.

$C_{res}$ for $I=3/2$ resonances are calculated by considering only the $\Delta I = 1$ portion of the current.  This term is proportional to $(e_u-e_d)$.  By substituting vector charges, $C_{res}$ for $I=3/2$ resonances is found to be $(1+Q_{W}^{p,LO})$.

The Christy-Bosted fit lies within $3\%$ of nearly all electromagnetic data points.  Our modification undoubtedly increases the uncertainty.  To be conservative we estimated our modifications increased the uncertainty to $10\%$.  

The Christy-Bosted fit also accounts for a smooth background.  To model the $\gamma Z$-box background we considered two limiting cases.  In the low x limit, the light quark distributions are expected to be equal and the corrective coefficient is
\be
\label{Cbackhigh}
C_{bkgd}|_{x \rightarrow 0}=\frac{\sum_{q=u,d,s} 2 e_q g^q_V f_q(x)}{\sum_{q=u,d,s}  (e_q)^2 f_q(x)}=1+Q_W^{p,LO}.
\ee
In the limit where there are only valence quarks 
\be
C_{bkgd}|_{\text{valence quarks}}=\frac{\sum_{q=u,u,d} 2 e_q g^q_V f_q(x)}{\sum_{q=u,u,d}  (e_q)^2 f_q(x)}=\frac{2}{3}+Q_W^{p,LO}.
\ee
We used these limits as error bounds and their average as the background correction.  Approximately half of the total contribution to $\Re\square_{\gamma Z}^V$ from the Christy-Bosted fit is due to this background modification. 

$F_2^{\gamma \gamma}$ is related to $F_1^{\gamma \gamma}$ by
\be
\label{f2}
F_2^{\gamma \gamma}=\frac{Q^2}{p\cdot q}\left(1+\frac{\sigma_L}{\sigma_T}\right)\frac{F_1^{\gamma \gamma}}{1+\frac{M^2Q^2}{(p\cdot q)^2}}.
\ee
We substituted $F_1^{\gamma Z}$ into the above expression to obtain $F_2^{\gamma Z}$.  We also assumed the modifications were the same for both the transverse and longitudinal cross sections.

Bosted and Christy~\cite{Bosted:2007xd} also have a fit for deuteron and neutron electromagnetic data which we used to modify the deuteron structure functions in~\cite{Carlson:2012yi}.  The corrective ratios for the deuteron resonances are listed in Table~\ref{amptable}.  Following the above analysis for the proton background, the limits to the deuteron background are $1+Q_W^{p,LO}$ and $4/5+Q_W^{p,LO}$.


\begin{table*}[htdp]

\begin{ruledtabular}

\caption{The seven Christy-Bosted resonances along with their electromagnetic helicity amplitudes along and corresponding corrective prefactors for both the proton and deuteron.  The ($pZ\rightarrow N_p^*$) helicity amplitudes are calculated by substituting $e_q\rightarrow g^q_V=T^3_q-2e_q\text{sin}^2 \theta_W$.  The ($n\gamma\rightarrow N_n^*$) and ($nZ\rightarrow N_n^*$) helicity amplitudes are calculated by exchanging $e_u \leftrightarrow e_d$ and $g^u_V \leftrightarrow g^d_A$, respectively, in the proton analysis.  The corrective prefactor for the background is also included.}
\begin{center}

\begin{tabular}{ll l ll}
resonance & proton electroproduction amplitudes & $C^p_{res}$ & $C^d_{res}$ \\ \hline
$P_{33}(1232)$	  & $A_{1/2}^{\gamma}\propto (e_u - e_d)$	 &                                         
        $1+Q_W^{p, LO}$ & $1+Q_W^{p, LO}$ \\
\\ 
$S_{11}(1535)$	&								
	$A_{1/2}^{\gamma} = \frac{1}{\sqrt{6}}
	\left( \sqrt{2}A_{10}\left(e_u - e_d \right)
	-  B_{10}  \left(\frac{5}{3}e_u+\frac{1}{3} e_d \right)\right)$     &
       		$\frac{1/3+2f_1}{1+2f_1}+Q_W^{p, LO}$ & 
$2\frac{(1+2f_1)(1/3+2f_1)}{(1+2f_1)^2+(1/3+2f_1)^2}+Q_W^{p, LO}$	\\
\\
\multirow{2}{*}{$D_{13}(1520)$}	&	 					 
	$A_{1/2}^{\gamma} = \frac{1}{\sqrt{6}}
	\left(   A_{10} \left(e_u - e_d \right)
	+ \sqrt{2} B_{10} \left(\frac{5}{3}e_u+\frac{1}{3} e_d \right)	\right)$         & 

\multirow{2}{*}{$\frac{(1-f_1)(1/3-f_1)+3f_1^2}{(1-f_1)^2+3f_1^2}+Q_W^{p,LO}$}   &

\multirow{2}{*}{$\frac{2(1-f_1)(1/3-f_1)+6f_1^2}{(1-f_1)^2+(1/3-f_1)^2+6f_1^2}+Q_W^{p,LO}$}
										\\
&	$A_{3/2}^{\gamma} = \frac{1}{\sqrt{2}} 
	A_{10} \left(e_u - e_d \right)$    &
                       		\\
\\
\multirow{2}{*}{$F_{15}(1680)$}	&	  					
	$A_{1/2}^{\gamma} = \sqrt{\frac{2}{5}} A_{20} \left(2e_u + e_d \right)+\sqrt{\frac{3}{5}} B_{20} 
	\left(\frac{4}{3}e_u  - \frac{1}{3} e_d \right) $     &

\multirow{2}{*}{$\frac{2/3(1-f_2)}{(1-f_2)^2+2f_2^2}+Q_W^{p,LO}$}    &

\multirow{2}{*}{$4\frac{1-f_2}{3(1-f_2)^2+6f_2^2+4/3}+Q_W^{p,LO}$}
											\\
&	$A_{3/2}^{\gamma} = 
 	\frac{2}{\sqrt{5}} A_{20} \left(2e_u + e_d \right)$	& \\
\\
$S_{11}(1650)$		&							
$A_{1/2}^{\gamma} = - \sqrt{\frac{2}{27}} B_{10} 
	\left( e_u  + 2 e_d \right)$		&		$\frac{1/3+2f_1}{1+2f_1}+Q_W^{p, LO}$	  & 
$2\frac{(1+2f_1)(1/3+2f_1)}{(1+2f_1)^2+(1/3+2f_1)^2}+Q_W^{p, LO}$	\\
\\
$P_{11}(1440)$		&						
$A_{1/2}^{\gamma} =  B_{00} 
	\left(\frac{4}{3}e_u  - \frac{1}{3} e_d \right)$	& $2/3+Q_W^{p, LO}$ &
$12/13+Q_W^{p, LO}$	\\
\\
$F_{37}(1950)$	&					
	$A_{1/2}^{\gamma}\propto (e_u-e_d)$ & $1+Q_W^{p, LO}$ & $1+Q_W^{p, LO}$ 
\\ \\
Background   &  & $\frac{5}{6}+Q_W^{p,LO}$ & $\frac{9}{10}+Q_W^{p,LO}$
\end{tabular}

\end{center}
\label{amptable}

\end{ruledtabular}

\end{table*}%


\section{Alternative Modification of $\bf{F_{1,2}^{\gamma \gamma}(x,Q^2)\rightarrow F_{1,2}^{\gamma Z}(x,Q^2)}$ in the Resonance Region}				\label{sec:alt}

The corrective prefactors for the Christy-Bosted fit can be modeled using a different technique.  The vector contribution to the $Z$-boson transition amplitudes can be isospin rotated into a sum of electromagnetic transition amplitudes, $p\gamma\rightarrow N^*_p$ and $n\gamma\rightarrow N^*_n$.  Neglecting strange quark contributions, these amplitudes are
\be
\label{Jemp}
\langle N^*_p|J_\mu^{\gamma (Z,V)}|p\rangle=e_u(g^u_V)\langle N^*_p|\bar{u} \gamma_\mu u|p\rangle+e_d(g^d_V)\langle N^*_p|\bar{d} \gamma_\mu d|p\rangle
\ee
and
\be
\label{Jemn}
\langle N^*_n|J_\mu^\gamma|n\rangle=e_u\langle N^*_n|\bar{u} \gamma_\mu u|n\rangle+e_d\langle N^*_n|\bar{d} \gamma_\mu d|n\rangle.
\ee
After performing an isopin rotation the neutron amplitude becomes
\be
\label{Jemn2}
\langle N^*_n|J_\mu^\gamma|n\rangle=e_u\langle N^*_p|\bar{d} \gamma_\mu d|p\rangle+e_d\langle N^*_p|\bar{u} \gamma_\mu u|p\rangle.
\ee
Further algebra on these amplitudes reveals
\be
\label{Jz}
\langle N^*_p|J_\mu^{Z,V}|p\rangle=\frac{1}{2}(1-4\text{sin}^2 \theta_W(0))\langle N^*_p|J_\mu^\gamma|p\rangle-\frac{1}{2}\langle N^*_n|J_\mu^\gamma|n\rangle.
\ee
$C_{res}$ can now be written as
\begin{align}
\label{Cresalt}
C_{res}&=Q_W^{p,LO}-\frac{\sum_\lambda A_{\lambda}^{\gamma,p}A_{\lambda}^{\gamma,n}}{\sum_\lambda (A_{\lambda}^{\gamma,p})^2}
\end{align}
Here, $p$ and $n$ identify the nucleon as a proton or neutron, respectively.  Gorchtein \textit{et al.}~\cite{Gorchtein:2011mz} constructed their $C_{res}$ expressions using photoproduction amplitudes listed in the Particle Data Group~\cite{Nakamura:2010zzi}.  Thus, their corrective prefactors lack $Q^2$-dependence.  To account for the amplitudes' $Q^2$-dependence, fits from MAID~\cite{Tiator:2011pw} can also be used.  

\begin{figure}
\begin{center}
\includegraphics[width = 3.37 in]{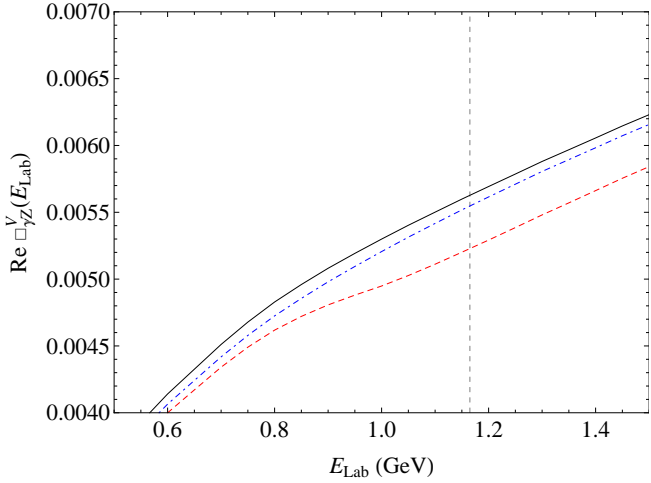}
\caption{$\Re\square_{\gamma Z}^V$ as a function of incoming electron energy.  The black curve is the result from our previous work and uses helicity amplitudes given by the quark model.  The blue, dot dashed curve is the result using corrective ratios from the PDG.  The red, dashed line is the result from using corrective ratios constructed with MAID helicity amplitudes.  The dashed, vertical line indicates the energy of the $Q_{weak}$ experiment.  All three models use the same modifications for isospin $3/2$ resonances and the smooth background.}
\label{plots}
\end{center}
\end{figure}

Fig.~\ref{plots} shows $\Re\square_{\gamma Z}^V$ calculated using both the quark model and MAID treatments of the structure functions.  Better agreement between MAID and the quark model was naively expected as the MAID fits were used to parameterize $\Lambda_{1,2}^2$.  The overall smaller value for $\Re\square_{\gamma Z}^V$ calculated by MAID is almost entirely due to the Roper resonance.  For the Roper, the quark model calculates a constant corrective prefactors while the MAID ratio rapidly approaches $Q_W^{p,LO}$ as $Q^2$ increases. The differences in the Roper resonance corrective prefactors were also the primary cause for the different deuteron asymmetry predictions in~\cite{Carlson:2012yi}.

Another notable feature of Fig.~\ref{plots} is that $\Re\square_{\gamma Z}^V$ hardly changes when the corrective ratios are calculated using PDG photoproduction amplitudes in place of the $Q^2$-dependent quark model.  $\Re\square_{\gamma Z}^V$ calculated using the quark model also remains relatively unchanged when using different values for $\Lambda_{1,2}^2$ values.  Both features are due to low $Q^2$ values dominating the integral.  Indeed, an analysis of the integral indicates that the mean $Q^2$ value is 0.4 GeV$^2$.  In applications with higher $Q^2$, such as the calculation of the deuteron asymmetry in~\cite{Carlson:2012yi}, the quark model and photoproduction corrective prefactors give quite different values.

It is important to note that Gorchtein \textit{et al.}~\cite{Gorchtein:2011mz} do not use the Christy-Bosted background in their analysis.  For the background they instead use the average of two Generalized Vector Dominance (GVD) models~\cite{Alwall:2004wk,Cvetic:2001ie}, isospin rotated for application to the $\gamma Z$-box and extrapolated down to the resonance region.  This averaging is the largest source of uncertainty for the Gorchtein \textit{et al.} calculation.  Recently, it has been claimed that this background uncertainty has been overestimated~\cite{Hall:2013hta}.


\section{Modification of Structure Function $\bf{F_{3}^{\gamma \gamma}(x,Q^2)\rightarrow F_{3}^{\gamma Z}(x,Q^2)}$ and the calculation of $\bf{\Re\square_{\gamma Z}^A}$}


Blunden \textit{et al.}~\cite{Blunden:2011rd} split their $\Re\square_{\gamma Z}^A$ analysis into elastic ($W^2=M^2$), resonance ($W_\pi^2\le$ $W^2\le 4$ GeV$^2$), and deep inelastic scaling ($W^2>4$ GeV$^2$) regions.  To allow for an easier comparison between our analysis and theirs, we used the same energy regions.  

As previously mentioned, the average $Q^2$ value within the $\Re\square_{\gamma Z}^V$ integral is about 0.4 GeV$^2$.  In contrast, the average $Q^2$ value within the $\Re\square_{\gamma Z}^A$ integral is about 80 GeV$^2$.  Thus, the axial contribution to the $\gamma Z$-box diagram is less sensitive to the modifications of the structure functions in the resonance region.  Because the axial box integral, Eq.(\ref{imdeltaax}), receives strong support from high $Q^2$, we follow the example of Blunden \textit{et al.} and use one loop running values of $\alpha(Q^2)$ and $\sin^2\theta_W(Q^2)$ in its evaluation.  Both running values are calculated in the $\overline{MS}$ renormalization scheme.

In the scaling region $F_3^{\gamma Z}$ can be directly calculated using parton distribution functions
\be
\label{f3dis}
F_3^{\gamma Z}(x,Q^2)=\sum_q2e_q g^q_A\left(q(x,Q^2)-\bar q(x,Q^2)\right).
\ee
Blunden \textit{et al.} use PDFs from ~\cite{Martin:2009iq}.  We chose PDFs given by CTEQ ~\cite{Lai:2010vv}.  CTEQ's uncertainty for the up quark is about 5\% and 10\% for the down quark.  To once again be conservative, we considered a 10\% uncertainty for this fit.

For $Q^2<1$ GeV$^2$ and $W^2>4$, we used the Model I modification to the PDFs discussed by Blunden \textit{et al.}, with $\Lambda^2=0.7$ GeV$^2$ and $Q_0^2=1$ GeV$^2$.  Blunden \textit{et al.} found an uncertainty of 10\% in this fit by varying $\Lambda^2$ within a reasonable range.  For the elastic contribution, we also follow the technique used by Blunden \textit{et al.}~\cite{Blunden:2011rd}.

The most significant departure from the Blunden \textit{et al.} analysis is in the resonance region.  In this region Blunden \textit{et al.} constructed $F_3^{\gamma Z}$ using axial current parameters of Lalakulich \textit{et al.}~\cite{Lalakulich:2006sw}.  Lalakulich \textit{et al.} obtained their parameters through a PCAC analysis of pionic decays of baryons.  Their fit accounts for four resonances but makes no attempt at estimating a smooth background, defering the determination of its form to future experiments.  As an aside, Lattice QCD calculations have reached a sufficient level of accuracy to calculate axial form factors \cite{Alexandrou:2010uk,Alexandrou:2010hf}.

Instead of repeating the Blunden \textit{et al.} resonance region analysis, we constructed $F_3^{\gamma Z}$ by once again modifying the Christy-Bosted fit.  Not only does this modification provide a smooth background, it also accounts for three more resonances.
In our analysis of the resonance region we repeated the technique outlined in Sec.~\ref{sec:f1f2mod}.  In the non-relativistic limit, $|\vec{k}|<<m_q$, the axial current becomes
\be
\label{axialcurrent}
\bar u(k',\lambda')\epsilon_+ \cdot \gamma\gamma_5 u(k,s')=\sqrt{2}\xi^\dagger_{\lambda'}S_+\xi_{s'}.
\ee
Continuing the use of the parameters in Sec.~\ref{sec:f1f2mod}, $F_3^{\gamma Z}$ can be expressed as
\begin{align}
F_3^{\gamma Z}&\Big|_{N\rightarrow res} = 
\nonumber\\
& 3\frac{2\nu}{q_z}\langle \psi_N\phi_N\chi_s\big|\big(2g_A^{q (3)}\big)\bigg[\frac{2m_q}{q_z}BS_+\bigg]^\dagger\big|\psi_{res}\phi_{res}\chi_\lambda\rangle
\nonumber\\
&\times 3\langle \psi_{res}\phi_{res}\chi_\lambda\big|e_q^{(3)}[AL_++BS_+]\big|\psi_N\phi_N\chi_s\rangle,
\end{align}
where $\nu$ is the energy of the exchanged boson.  For our calculation we took the mass of the struck quark $m_q$ to be 0.3 GeV.  Table~\ref{amptableax} summarizes the corrective prefactors to obtain $F_3^{\gamma Z}$.  As with the corrective prefactors for $F_{1,2}^{\gamma Z}$, we estimate the uncertainty of the $F_3^{\gamma Z}$ prefactors to be 10\%.

The smooth background is once again modified by taking the low x and valence quark limits.  For low x, a quark and anti-quark are equally likely to be struck.  Thus,
\be
\label{Cbackhigh2}
C_{bkgd}|_{x \rightarrow 0}=\frac{\sum_{q=u,d,s} 2 e_q g^q_A f_q(x)}{\frac{1}{2}\sum_{q=u,d,s}  (e_q)^2 f_q(x)}=0.
\ee
In the limit where valence quarks are equally likely to be struck
\be
C_{bkgd}|_{\text{valence quarks}}=\frac{\sum_{q=u,u,d} 2 e_q g^q_A f_q(x)}{\frac{1}{2}\sum_{q=u,u,d}  (e_q)^2 f_q(x)}=\frac{10}{3}.
\ee
These limits were taken as the uncertainty bounds and their average as the modification for the smooth background.

We also calculated $F_3^{\gamma Z}$ for the deuteron in~\cite{Carlson:2012yi}.  The corrective ratios for the deuteron resonances are listed in Table~\ref{amptableax}.  Following the above analysis for the proton background, the limits to the deuteron background are $0$ and $18/5$.

\begin{figure}
\begin{center}
\includegraphics[width = 3.37 in]{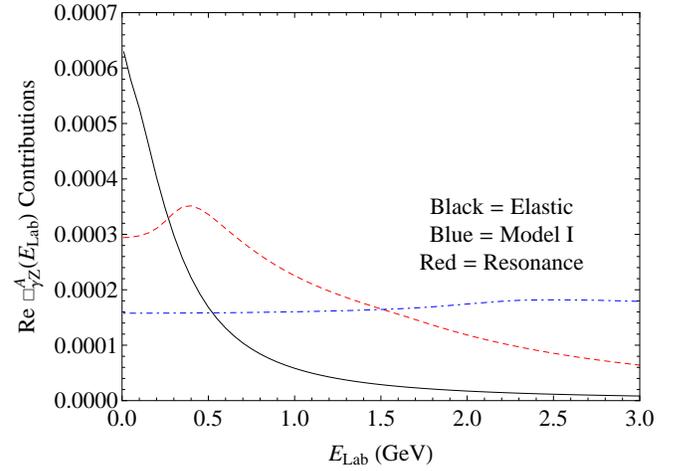}
\caption{Elastic (black, solid curve), resonance (red, dashed curve), and model I (blue, dot dashed curve) contributions to the axial box.}
\label{axialboxsplit}
\end{center}
\end{figure}

\begin{figure}
\begin{center}
\includegraphics[width = 3.37 in]{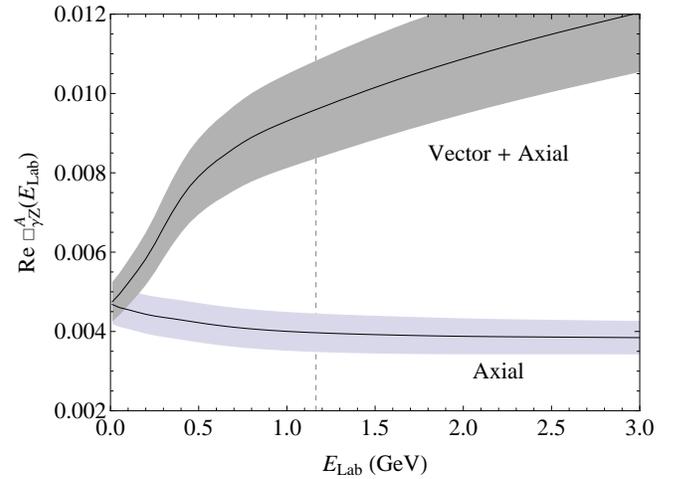}
\caption{The axial box.  We also add the axial box to our previous vector calculation~\cite{Rislow:2010vi} to obtain the total box.  The dashed, vertical line indicates the energy of the $Q_{weak}$ experiment.  }
\label{axialbox}
\end{center}
\end{figure}

Figs.~\ref{axialboxsplit} and ~\ref{axialbox} display the results for $\Re\square_{\gamma Z}^A$.  As can be seen, the scaling region dominates.  At the $Q_{weak}$ energy, $\Re\square_{\gamma Z}^A=0.0040\pm 0.0005$.


\begin{table*}[htdp]

\begin{ruledtabular}

\caption{The seven Christy-Bosted resonances along with their axial helicity amplitudes and corrective prefactors for both the proton and deuteron.  The neutron amplitude is calculated by exchanging $g^u_A \leftrightarrow g^d_A$.}
\begin{center}

\begin{tabular}{l l l l}
resonance & proton axial current amplitudes & $C_{res}^p$ & $C_{res}^d$\\ \hline
$P_{33}(1232)$	  & $A_{1/2}^{Z,A}\propto (g^u_A - g^d_A)\frac{4m_q\nu}{q_z^2}$	 &                                         
        $2\frac{4m_q\nu}{q_z^2}$ & $2\frac{4m_q\nu}{q_z^2}$\\
\\
$S_{11}(1535)$	&								
	$A_{1/2}^{Z,A} = -\frac{1}{\sqrt{6}}
	B_{10}\left(\frac{5}{3}g^u_A+\frac{1}{3}g^d_A \right)\frac{4m_q\nu}{q_z^2}$      &
       		$\frac{1}{3(2f_1+1)}\frac{16m_q\nu}{3q_z^2}$ & $\frac{(1+2f_1)+(1/3+2f_1)}{(1+2f_1)^2+(1/3+2f_1)^2}\frac{16m_q\nu}{3q_z^2}$	\\
\\
\multirow{2}{*}{$D_{13}(1520)$}	&	 					 
	$A_{1/2}^{Z,A} = \sqrt{\frac{2}{6}}
	  B_{10} \left(\frac{5}{3}g^u_A+\frac{1}{3}g^d_A \right)\frac{4m_q\nu}{q_z^2}$         & 

\multirow{2}{*}{$\frac{1-f_1}{(f_1-1)^2+3f_1^2}\frac{16m_q\nu}{3q_z^2}$} & \multirow{2}{*}{$\frac{(1-f_1)-(f_1-1/3)}{(1-f_1)^2+(f_1-1/3)^2+6f_1^1}\frac{16m_q\nu}{3q_z^2}$}
										\\
&	$A_{3/2}^{Z,A} = 0$    & &
                       		\\
\\
\multirow{2}{*}{$F_{15}(1680)$}	&	  					
	$A_{1/2}^{Z,A} = \sqrt{\frac{3}{5}} B_{20} 
	\left(\frac{4}{3}g^u_A  - \frac{1}{3} g^d_A \right)\frac{4m_q\nu}{q_z^2}$     &

\multirow{2}{*}{$\frac{(1-f_2)}{(1-f_2)^2+2f_2^2}\frac{20m_q\nu}{3q_z^2}$} & \multirow{2}{*}{$\frac{(1-f_2)+2/3}{(1-f_2)^2+2f_2^2+4/9}\frac{20m_q\nu}{3q_z^2}$}
											\\
&	$A_{3/2}^{Z,A} = 0$	& & \\

\\
$S_{11}(1650)$		&							
$A_{1/2}^\gamma = -\sqrt{\frac{2}{27}} B_{10} 
	\left(g^u_A  + 2g^d_A \right)\frac{4m_q\nu}{q_z^2}$		&		$\frac{1}{3(2f_1+1)}\frac{16m_q\nu}{3q_z^2}$ & $\frac{(1+2f_1)+(1/3+2f_1)}{(1+2f_1)^2+(1/3+2f_1)^2}\frac{16m_q\nu}{3q_z^2}$		\\
\\
$P_{11}(1440)$		&						
$A_{1/2}^{Z,A} =  B_{00} 
	\left(\frac{4}{3}g^u_A  - \frac{1}{3} g^d_A \right)\frac{4m_q\nu}{q_z^2}$	& $\frac{20m_q\nu}{3q_z^2}$ & $\frac{100m_q\nu}{13q_z^2}$	\\
\\
$F_{37}(1950)$	&					
	$A_{1/2}^{Z,A}\propto (g^u_A - g^d_A\frac{4m_q\nu}{q_z^2}$ & $2\frac{4m_q\nu}{q_z^2}$ & $2\frac{4m_q\nu}{q_z^2}$
\\ \\
Background   &  & $\frac{5}{3}$ & $\frac{9}{5}$
\end{tabular}

\end{center}
\label{amptableax}

\end{ruledtabular}

\end{table*}%



\section{Conclusions}


Adding the axial box to our original vector box calculation~\cite{Rislow:2010vi}, our constituent quark model yields a total $\gamma Z$-box value of
\begin{align}
\Re\square_{\gamma Z}(E=1.165\text{ GeV})|_{\text{total}}=(9.7\pm 1.4)\times 10^{-3}.
\end{align}
The errors from both the axial and vector calculations were added directly.  If added in quadrature, the uncertainty reduces to $1\times 10^{-3}$.

The total $\gamma Z$-box value from Blunden \textit{et al.}~\cite{Blunden:2011rd} is
\begin{align}
\Re\square_{\gamma Z}(E=1.165\text{ GeV})|_{\text{total}}=(8.4^{+1.1}_{-0.6})\times 10^{-3}.
\end{align}

These two calculations are in agreement within uncertainties.  Each calculation also has error bounds below the error budget of the $Q_{weak}$ experiment.  

The question remains which calculations the $Q_{weak}$ collaboration should use in their analysis.  The disagreement between the various calculations is largely due to the treatment of the $\gamma Z$ structure functions in the resonance region.  We believe the collaboration will be equally well-served by either $\Re\square^A_{\gamma Z}$ calculation.  $\Re\square^A_{\gamma Z}$ is not very sensitive to the resonance region modifications since its integrals get much of their support from high $Q^2$.  $F_3^{\gamma Z}$ in the scaling region can be constructed using fits to parton distribution data.

Which $\Re\square^V_{\gamma Z}$ calculation to use is more open to debate.  The vector integrals receive much of their support from the resonance region and are thus sensitive to the modification $F_{1,2}^{\gamma \gamma}\rightarrow F_{1,2}^{\gamma Z}$.  In Sec.~\ref{sec:alt} we showed that there is little difference between modifying the Christy-Bosted resonance fits using our constituent quark model~\cite{Rislow:2010vi} or photoproduction amplitudes from the Particle Data Group (as in~\cite{Gorchtein:2011mz}).  Differences arise between \cite{Rislow:2010vi} and \cite{Gorchtein:2011mz} because of the treatments of the resonance region background.  We continue modifying the Christy-Bosted background fit while Gorchtein \textit{et al.} modify two GVD fits to low $Q^2$, high $W^2$ data and extrapolate them down to the resonance region.  We believe our modification is more satsifactory since it does not involve any extrapolations.  We cannot comment on the vector calculation of~\cite{Sibirtsev:2010zg} since they provide few details of their model.


\begin{acknowledgments}

We thank Peter Blunden and Wally Melnitchouk for useful conversations and the National Science Foundation for support under Grant No.  PHY-1205905.

\end{acknowledgments}

  \vskip -15pt

\clearpage
\bibliography{structurefunctions}

\newpage

\end{document}